\documentclass[prl,aps,reprint,a4paper,superscriptaddress,citeautoscript,floatfix]{revtex4-2}
\usepackage{amsmath,amsfonts,amssymb}
\usepackage{newtxtext,newtxmath}

\usepackage[pdftex]{graphicx}
\usepackage{microtype}
\usepackage[svgnames]{xcolor}
\usepackage[colorlinks=True,linkcolor=DarkRed,citecolor=ForestGreen,urlcolor=MediumBlue,
	pdfstartview=FitH,bookmarks=False,pdfpagemode=UseNone]{hyperref}

\let\vec\mathbf
\usepackage{bm}
\usepackage{color}
\usepackage{booktabs}
\usepackage{subcaption}
\makeatletter
\DeclareCaptionJustification{leftjustified}{%
  \leftskip\z@skip\rightskip\z@skip\@rightskip\z@skip\parfillskip\@flushglue}
\makeatother
\makeatletter
\newwrite\SM@aux
\newcommand{\SMbibstart}{%
  \immediate\openout\SM@aux=\jobname-SM.aux
  \immediate\write\SM@aux{\relax}%
  \immediate\write\SM@aux{\string\citation{REVTEX42Control}}%
  \immediate\write\SM@aux{\string\citation{apsrev42Control}}%
  \immediate\write\SM@aux{\string\bibstyle{apsrev4-2}}%
  \immediate\write\SM@aux{\string\bibdata{\jobname Notes,biblio}}%
}
\DeclareRobustCommand{\citeSM}[1]{%
  \immediate\write\SM@aux{\string\citation{#1}}%
  \def\SM@keys{}%
  \@for\SM@k:=#1\do{\edef\SM@keys{\SM@keys\ifx\SM@keys\@empty\else,\fi SM-\SM@k}}%
  \futurelet\SM@tok\SM@decide
}
\def\SM@decide{%
  \let\SM@next\SM@citenopunct
  \ifx\SM@tok.\let\SM@next\SM@citepunct\fi
  \ifx\SM@tok,\let\SM@next\SM@citepunct\fi
  \ifx\SM@tok;\let\SM@next\SM@citepunct\fi
  \ifx\SM@tok:\let\SM@next\SM@citepunct\fi
  \SM@next}
\def\SM@citenopunct{\SM@citepunct{}}
\def\SM@citepunct#1{%
  \let\SM@savedfilesw\if@filesw \@fileswfalse
  \expandafter\cite\expandafter{\SM@keys}#1\let\if@filesw\SM@savedfilesw}
\newcommand{\SMbibsetup}{%
  \immediate\closeout\SM@aux
  \let\@FMN@list\@empty
  \let\SM@origthebib\thebibliography
  \def\thebibliography##1{\SM@origthebib{##1}\SM@patchbibitem}%
  \def\SM@patchbibitem{%
    \let\SM@origbibitem\bibitem
    \def\bibitem{\@ifnextchar[\SM@bibitem@opt\SM@bibitem@noopt}}%
  \def\SM@bibitem@opt[##1]##2{\SM@origbibitem[##1]{SM-##2}}%
  \def\SM@bibitem@noopt##1{\SM@origbibitem{SM-##1}}%
  \let\SM@origlabel\label
  \def\label##1{\SM@origlabel{SM-##1}}%
}
\makeatother

\def\<{\langle}
\def\>{\rangle}

\begin{document}
\title{How Twist Angle Inhomogeneity Masks the Superconducting Order Parameter Symmetry}

\author{Ilaria Maccari}

\email{imaccari@phys.ethz.ch}
\affiliation{Institute for Theoretical Physics, ETH Zurich, Wolfgang-Pauli-Strasse 27, 8093 Zurich, Switzerland}
\author{Louk Rademaker}
\affiliation{Department of Quantum Matter Physics, University of Geneva, 24 quai Ernest-Ansermet, 1211 Geneva, Switzerland}
\affiliation{Institute-Lorentz for Theoretical Physics, Leiden University, PO Box 9506, 2300 Leiden, The Netherlands}
\author{Giulia Venditti}
\email{giulia.venditti@unige.ch}
\affiliation{Department of Quantum Matter Physics, University of Geneva, 24 quai Ernest-Ansermet, 1211 Geneva, Switzerland}

\date{\today}

\begin{abstract}
Two-dimensional superconductors, including twisted multilayer graphene, should exhibit a BKT transition, and the $T$ dependence of the superfluid stiffness should distinguish between nodal or gapped order parameter symmetries. This picture dramatically changes when spatially correlated disorder is taken into account. Such correlations naturally arise in moiré systems due to twist angle inhomogeneities, which we model using elasticity theory. Using a random impedance network based on realistic disorder in the local $T_c$, we show that such long-ranged inhomogeneities not only smear the BKT transition, but they can obscure the distinction between nodal and fully gapped superconducting order parameters. We propose that the real part of the conductivity can be used as a key diagnostic observable to probe the relevance of correlated disorder.
\end{abstract}

\maketitle

Two-dimensional (2D) superconductivity has attracted growing interest over the last decade, driven by remarkable advances in materials fabrication that have made truly crystalline 2D superconductors experimentally accessible \cite{saito_highly_2016, qiu_recent_2021, ji_two-dimensional_2024}. These span a wide range of platforms, from oxide heterostructures such as SrTiO$_3$- and KTaO$_3$-based interfaces~\cite{reyren2007superconducting, liu_two-dimensional_2021, chen_electric_2021}, to atomically thin van der Waals superconductors such as NbSe$_2$~\cite{xi_ising_2016} and MoS$_2$~\cite{lu_evidence_2015}, and, most recently, a broad family of moiré systems~\cite{andrei_marvels_2021,Kennes2021} including magic-angle twisted bilayer graphene (tBG) \cite{Cao2018}, twisted trilayer graphene \cite{zhou_gate-tunable_2025}, twisted four- and five-layer graphene \cite{Park2022},
twisted transition-metal dichalcogenides bilayers \cite{wang_correlated_2020}, and rhombohedral multilayer graphene \cite{zhou_superconductivity_2021, han_signatures_2025}. 
These new classes of 2D materials offer an unprecedented opportunity to study the interplay of thermal and quantum fluctuations in highly tunable platforms.

In contrast to bulk systems, in 2D superconductors the nature of the superconducting (SC) transition is radically changed. While in three-dimensional (3D) superconductors the SC transition is in general governed by the formation of Cooper pairs that become immediately phase-coherent, in two dimensions this is no longer the case: there, Cooper pairs can form without phase coherence, which is only established at a lower temperature through a topological phase transition toward a quasi-long-range ordered state. This is the celebrated Berezinskii-Kosterlitz-Thouless (BKT) transition \cite{berezinsky1972, kosterlitz1973ordering, kosterlitz1974critical}, driven by the proliferation of free topological excitations of the SC phase, i.e. vortex-antivortex pairs. The BKT's most striking hallmark is the universal jump in the superfluid stiffness $J_s$ at the critical temperature which follows the Nelson-Kosterlitz \cite{nelson1977universal} relation: $J_s(T_{\rm BKT}^-)=2T_{\rm BKT}/\pi; \, J_s(T_{\rm BKT}^+)= 0$.

A second key aspect of 2D superconductivity is its interplay with disorder. At zero temperature, disorder is known to drive a superconductor-to-insulator quantum phase transition~\cite{wang_quantum_2023}.  At finite temperature, it can significantly modify the BKT signatures while leaving the universality class of the critical point intact, in accordance with the Harris criterion~\cite{harris_1974}.  In particular, many experiments have reported a smearing of the expected stiffness jump near $T_{\rm BKT}$~\cite{Fiory_superconducting_1983, Turneaure_thermalphase_2000,Zuev_correlation_ybco_2005, Broun_superfluid_2007,Sacepe_disorder_2008,Mondal_VortexCore_2011, Yong_Robustness2013,Mandal_destruction_2020, mallik_superfluid_2022, jarjour_superfluid_2023}. While this broadening was initially attributed to disorder strength~\cite{benfatto2009broadening}, it was later understood to be controlled by the spatial correlations of the disorder rather than its magnitude~\cite{maccari2017broadening, maccari2019disordered,Weitzel2023}. When spatial correlations become sufficiently long ranged, the effect can be even more substantial: superconductivity nucleates in isolated islands, and the system behaves as a Josephson junction array. 
 In this limit, as discussed for LaAlO$_3$/SrTiO$_3$ interfaces~\cite{caprara2011effective,caprara2013multiband,bucheli2013metal,caprara2014inhomogeneous,singh2018competition, bucheli2015pseudo, venditti2019nonlinear, venditti2020superfluid, venditti2021finite,Venditti2023RIN}, the phase transition loses its BKT character entirely, becoming a percolative transition among superconducting islands.

\begin{figure}
    \centering
    \includegraphics[width=\linewidth]{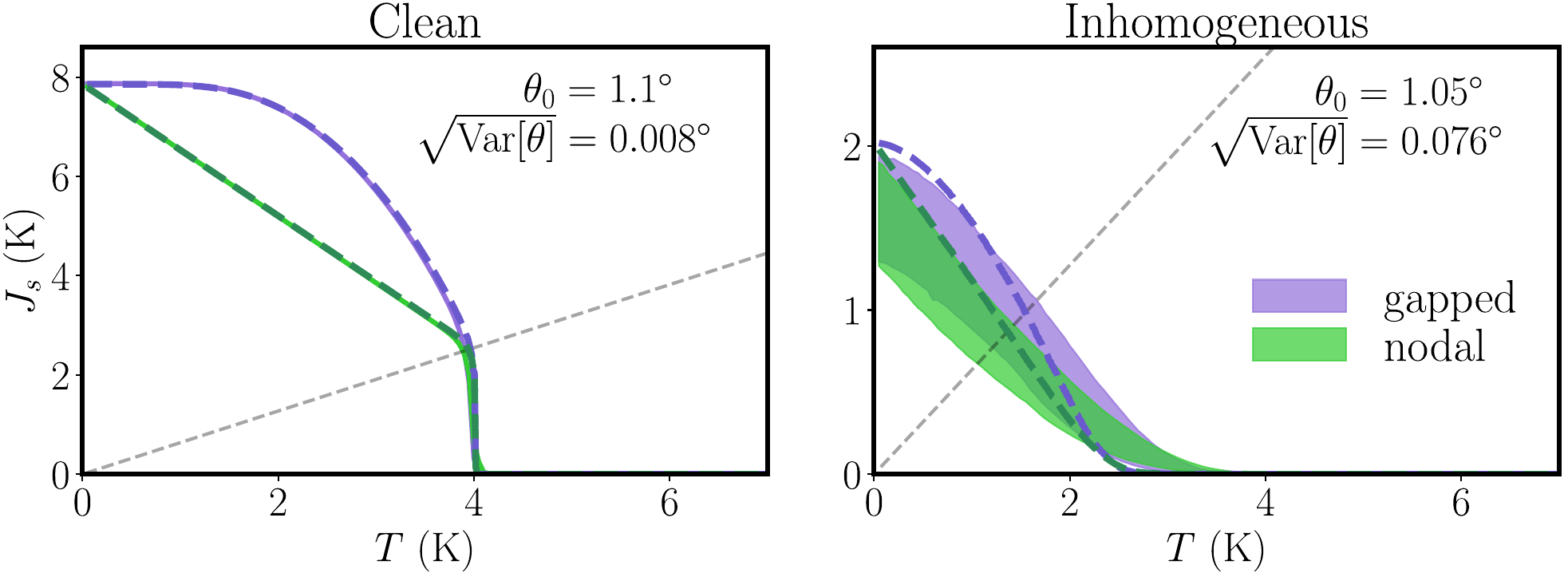}
    \caption{
    Comparison of the superfluid stiffness $J_s$ between gapped (violet) and nodal (green) superconducting order parameters in a clean (left) and long-range correlated inhomogeneous (right) system.
    Filled lines show the exact random impedance network (RIN) calculation, dashed lines are their effective medium theory solution (see main text).
    The dashed grey line is the universal BKT critical line $2T/\pi$.
    The width of the RIN solutions represents the distribution of disorder realizations. 
    Whereas in the clean case (left), there is a distinct difference between gapped and nodal superconductors, this difference disappears when correlated disorder is present (right).
    }
    \label{fig:BCS-vs-nodal}
\end{figure}
Moiré superconductors are particularly susceptible to this scenario  \cite{Beechem2014,rhodes_disorder_2019, Uri2020, Turkel2022, Grover2022,Park2022,Dolleman2024}.
In these systems, 
the critical temperature--as well as normal-state transport properties~\cite{Wilson2020,Padhi2020}-- are exquisitely sensitive to the local twist angle, so any spatial 
variation in the stacking configuration maps directly onto an inhomogeneous landscape of 
local $T_c$'s. 
Indeed, a broadening of the BKT stiffness jump has recently been reported 
in moiré systems, including twisted bilayer~\cite{Tanaka2025} and trilayer 
graphene~\cite{Banerjee2025, Mukherjee2025, Park2025}.

Away from the transition, the low-temperature behavior of the superfluid stiffness has been used as a probe of order-parameter symmetry. For a fully gapped $s$-wave superconductor, the stiffness is nearly constant at low temperatures, $J_s(T) \approx J_s(0)\left(1 - a\,e^{-\Delta/T}\right)$, whereas in a clean nodal superconductor it decays linearly, $J_s(T) \approx J_s(0)\left(1 - a\,\frac{T}{\Delta}\right)$, see Fig.~\ref{fig:BCS-vs-nodal}, left.
The observed linear-in-$T$ decay of the stiffness in twisted graphene systems has been cited as evidence for nodal superconductivity~\cite{Banerjee2025, Tanaka2025, Mukherjee2025, Park2025}.

In this Letter, we demonstrate that correlations in twist angle disorder are an intrinsic feature of moiré systems, naturally emerging from the solid's elasticity. Such large-scale inhomogeneities not only broaden the BKT transition but can also generically produce a power-law suppression of the superfluid stiffness at low temperatures, masking the underlying order-parameter symmetry. We show this in Fig.~\ref{fig:BCS-vs-nodal}, right: with correlated disorder, one cannot tell gapped from nodal superconductors. 
Our results call for a reinterpretation of existing experimental data and highlight the need for probes that can disentangle intrinsic pairing symmetry from disorder-induced renormalization of the superfluid response.

{\em Long-range correlated disorder---}Disorder in crystalline materials 
arises from a variety of microscopic sources, such as  vacancies, interstitials, chemical substitutions, and so forth. On a coarse-grained scale, these can be modeled as local random forces $\vec{f} (\vec{x})$ acting on the medium, {which} in turn {produce} local displacements $\vec{u}(\vec{x})$ {governed by} 
\begin{equation}
	\Delta \vec{u}(\vec{x}) + \frac{1}{1 - 2 \nu} \vec{\nabla} (\vec{\nabla} \cdot \vec{u}(\vec{x}) ) = -\rho \vec{f}(\vec{x}) \frac{2 (1 + \nu)}{E},
\label{Eq:ForcesDisplacement}
\end{equation}
where $\nu$ is Poisson's ratio, $E$ is Young's modulus {, and $\rho$ is the medium density~\cite{landau1986theory}}. {Being linear, Eq.~\eqref{Eq:ForcesDisplacement} is solved by 
Green's functions, $ \vec{u}(\vec{x}) = \int d^2 x' \hat{G}(\vec{x} - \vec{x'}) \vec{F}(\vec{x})$, where $\vec{F}(\vec{x}) = \rho \vec{f}(\vec{x}) \frac{2 (1 + \nu)}{E}$}. In 2D, the relevant components of the elastic Green's tensor $\hat{G}(\vec{x} - \vec{x'})$ read 
\begin{align}
	G_{xx} (\vec{x}) & = 
		- \frac{3-4\nu}{16 \pi(1-\nu)} \left( 2 \gamma + \log |\vec{x}|^2 \right)
		+ \frac{x^2}{8 \pi |\vec{x}|^2 (1 - \nu)};
        \label{Eq:GF1}
		\\
	G_{xy} (\vec{x}) & = 
        \frac{xy}{8 \pi |\vec{x}|^2 (1 - \nu)}.
    \label{Eq:GF2}
\end{align}
Importantly, even perfectly uncorrelated random forces,
\begin{equation}
	\langle f_a(\vec{x}_1') f_b(\vec{x}_2') \rangle \propto \delta_{ab} \delta(\vec{x}_1'-\vec{x}_2'), \qquad a,b=x,y
\end{equation}
give rise to long-range correlated displacements as a consequence of the nonlocal nature of the elastic Green's tensor.

\begin{figure}
    \centering
    \includegraphics[width=\linewidth]{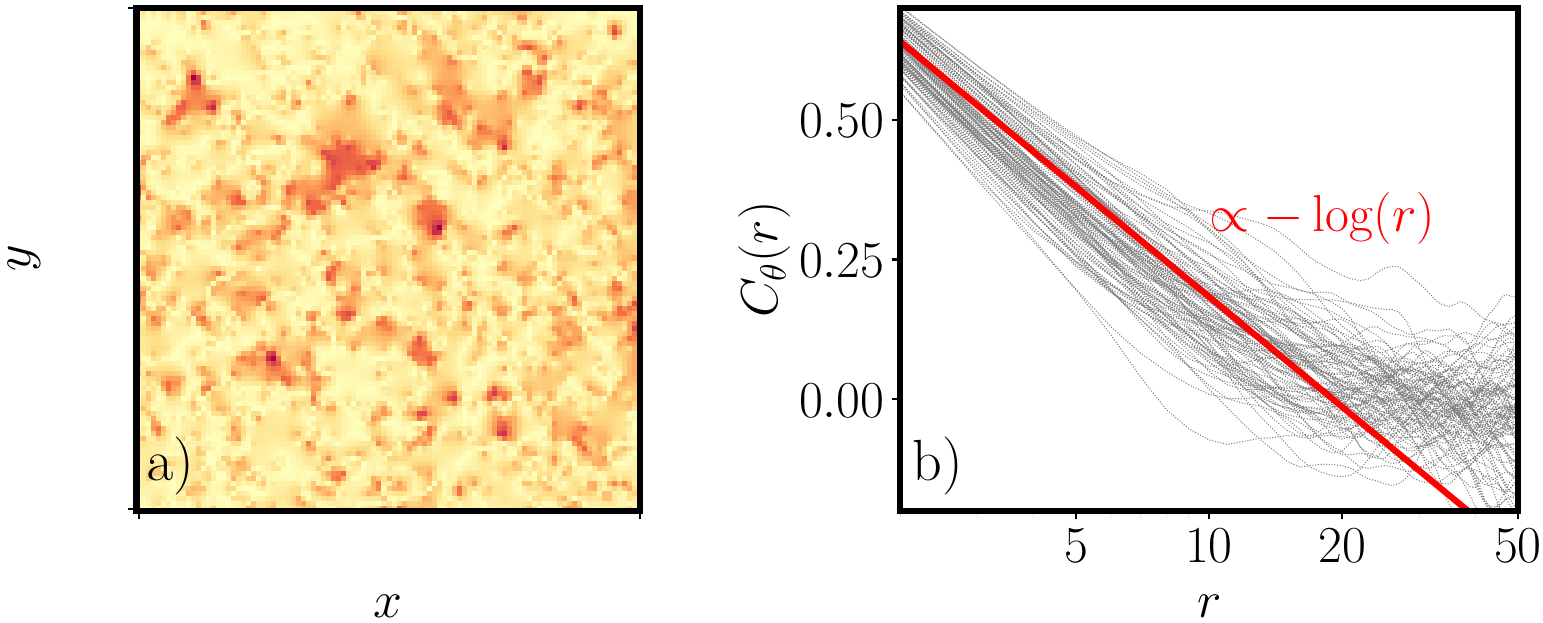}
    \caption{(a) Typical twist angle map, created by numerically solving the elasticity equations, given a random impurity density of $n_{\rm imp}=0.1$ on a $100\times 100$ lattice. The magnitude of the twist angle variations is in arbitrary units. (b) The twist angle inhomogeneity correlation function $C_\theta(\vec{r})$ is long ranged. Gray dashed lines are correlation functions for 100 different realizations like the one in (a); the red solid line is a logarithm following Eq.~\eqref{Eq:CorrelationFunctionLogSolution}.}
    \label{fig:twistanglemaps}
\end{figure}

In moiré systems, local lattice displacements induce variations in the local twist angle $\delta \theta(\vec{x})$ \cite{Balents2019}:
\begin{align}
    \delta \theta(\vec{x}) = \hat{z} \cdot \left[ \vec{\nabla} \times \vec{u}(\vec{x}) \right].
\label{Eq:ThetaField}
\end{align}
The typical twist-angle inhomogeneity is captured by the correlation function
\begin{equation}
	C_\theta (\vec{x}_1 - \vec{x}_2) = \langle \delta\theta(\vec{x}_1) \delta \theta(\vec{x}_2) \rangle.
    \label{Eq:TwistAngleCorrelationFunction}
\end{equation}
{Using Eq.~\eqref{Eq:GF1}-\eqref{Eq:GF2} and expressing the curl in 
Eq.~\eqref{Eq:ThetaField} in terms of the antisymmetric tensor $\hat{\epsilon}$, 
the correlation function can be evaluated in Fourier space,}
\begin{equation}
\begin{split}
    C_\theta  (\vec{r})
	&\propto \sum_{\vec{k}} \vec{k} \cdot \hat{\epsilon}  \cdot \hat{G}_{\vec k}  \cdot \hat{G}^\intercal_{-\vec{k}}  \cdot \hat{\epsilon}^{\intercal} \cdot \vec{k}
	\; \; e^{-i \vec{k} \cdot \vec{r}} =\\
    & = \sum_{\vec{k}} \frac{1}{|\vec{k}|^2} \; e^{-i \vec{k} \cdot \vec{r}}  \propto {\rm constant} - \log |\vec{r}|.
    \end{split}
    \label{Eq:CorrelationFunctionLogSolution}
\end{equation}
{Regardless of the microscopic details of the disorder,} the surprisingly simple result is, therefore, that elasticity implies a twist-angle inhomogeneity that is logarithmically long-range correlated, even for a homeopathic amount of impurities.
To obtain specific realizations, we discretize Eq.~\eqref{Eq:ForcesDisplacement} on a square lattice of linear size $L$ with open boundary conditions, taking $\nu=0.19$, within the range $\nu\approx 0.16-0.2$ reported for graphene \cite{Bi2019,Lebedeva2019,Kang2025}. 
Random forces act on a fraction $n_{\rm imp}$ of the coarse-grained lattice sites. Such {\em coarse-grained impurities} can originate from sample defects, local strain, substrate impurities, and intrinsic elastic relaxation, and their strength $|{\bf F}|$ is difficult to quantify.
Fig.~\ref{fig:twistanglemaps}(a) shows a typical twist-angle map for $n_{\rm imp}=0.1$ and $L=100$, qualitatively similar to those observed in twisted bilayer and trilayer graphene \cite{Uri2020,Turkel2022,Beechem2014,Hu.2024.CPL}. 
Fig.~\ref{fig:twistanglemaps}(b) shows that the numerically computed $C_\theta$ is consistent with logarithmic decay.
The logarithmic behavior is constant as a function of $n_{\rm imp}F^2$, whereas the variance of $\delta\theta$ scales proportionally with it~\cite{Supplementary}.
Thus, the relevant experimentally accessible quantity is the variance of twist angle inhomogeneities, Var[$\theta$], which we take as our tuning knob by rescaling the $\delta\theta$ maps to be comparable with experimental observations~\cite{Uri2020,Beechem2014,Turkel2022,Hu.2024.CPL}, while keeping $|{\bf F}|\in(0,1]$ fixed. 
At fixed Var[$\theta$], the transport properties are governed by the long-range correlations rather than the value of $n_{\rm imp}$~\cite{Supplementary}, or the system size $L$. Hence, hereafter, we will present our results for $n_{\rm imp}=0.1$ and $L=100$.

The long-range correlations we consider are generic to any disorder that couples to the lattice via strain. 
We argue that this elastic channel generically injects long-range correlations which will dominate the transition and the superfluid properties, and could naturally extend to other electronic properties. 
Any other short-range or pointlike disorder will be averaged out. 
Note that we are neglecting lattice defects, dislocations, and disclinations. Due to their topological nature, these defects are highly nonlocal as well, and consequently would only add to the long-range nature of disorder. 

{\em From disorder to transport---}Having established that twist-angle 
inhomogeneities are both inevitable and logarithmically long-range correlated, 
we now turn to their observable consequences for the superconducting transition. We focus 
on tBG as a paradigmatic moiré superconductor. 
Our approach is, however, general and applies to the wider family of moiré systems, particularly twisted multilayer graphene.

{The superfluid stiffness is experimentally accessible through microwave resonance 
measurements \cite{singh2018competition,mallik_superfluid_2022,Venditti2023RIN,Weitzel2023,Banerjee2025,Tanaka2025}, which probe the imaginary part $\sigma_2$ of the optical conductivity 
$\sigma(T, \omega_0) = \sigma_1(T, \omega_0) - i\sigma_2(T, \omega_0)$ at a resonance frequency 
$\omega_0$ below the gap, yielding
\begin{equation}
    J_s(T, \omega_0) = \frac{\hbar^2\omega_0}{4e^2}\,\sigma_2(T, \omega_0).
    \label{Eq:RelationSuperfluidStiffnesstoConductivity}
\end{equation}}

The disorder patterns of Fig.~\ref{fig:twistanglemaps} can be viewed as a coarse-grained lattice, where the local twist angle at the site $i$, i.e., $\theta_i=\theta_0+r_{\rm maps}\delta\theta_i$ with $r_{\rm maps}$ tuning Var[$\theta$], translates into a local critical temperature via the phenomenological relation
\begin{equation}
    T_{c,i} = T_{c, {\rm max}} 
    \left( 1 - \frac{(\theta_i - \theta_{\rm MA})^2}{\Delta \theta^2}\right)
    \label{eq:Tc_phenomen}
\end{equation}
where $\theta_{\rm MA}$ is the magic angle where superconductivity has the highest $T_c$, and $\Delta \theta$ is the width of the window of twist angles where there is superconductivity. For tBG, we adopt $T_{c, {\rm max}} \approx 4$\,K, $\Delta \theta \approx 0.1^\circ$, and the magic angle $\theta_{\rm MA} \approx 1.1^\circ$, based on experimental studies \cite{Saito2020,Koblischka.2024,Supplementary}.

We model the resistive and superfluid response of the corresponding disordered 
superconductor by means of a random impedance network (RIN)~\cite{Venditti2023RIN}, 
in which the disordered landscape of local $T_c$'s is realized through the bonds: 
each bond $i$ represents a mesoscopic superconducting island characterized by its 
local BKT transition temperature $T_{c,i}$.
Typical sizes of superconducting puddles  are of the order of $\sim 0.1$ to $1\mu$m in tBG \cite{Beechem2014,Uri2020,Dolleman2024,Rothstein2026}.
Throughout, we focus on a nodeless $s$-wave 
order parameter, showing how long-range correlations alone can modify its low-temperature behavior.
Each bond carries a local impedance
\begin{equation}
    Z_i(T,\omega_0) = R_i(T) + i\omega_0 L_i(T),
    \label{eq:local_impedance}
\end{equation}
where, different from the RIN model studied in~\cite{Venditti2023RIN}, the resistive and inductive components now encode both the amplitude and phase fluctuations of the local superconducting order parameter as a function of the temperature.

For $T < T_{c,i}$, the bond is superconducting: $R_i(T) = 0$ and the inductive response is governed by the local superfluid stiffness. For a nodeless $s$-wave gap, the normalized stiffness can be written as
\begin{equation}
    \frac{J_{s,i}(T)}{J_{s,i}(0)} = f_{\rm BCS}(T,T_{c,i}^0) =  g(\tau_i)
    \tanh\!\left(\frac{1.764\,g(\tau_i)}{2\tau_i}\right);
    \label{eq:Js_local}
\end{equation}
where  $\tau_i \equiv T/T_{c,i}^0$ and $g(\tau_i) = \sqrt{\cos\!\left(\frac{\pi}{2}\tau_i^2\right)}$. The local mean-field critical temperature $T_{c,i}^0$ is fixed, given $T_{c,i}$, by the universal-jump condition $J_{s,i}^0(T_{c,i}) = 2T_{c,i}/\pi$. Finally, the local kinetic inductance is related to the stiffness via
\begin{equation}
    L_{i}(T) = \frac{\hbar^2}{4e^2 k_B J_{s,i}(T)},
    \label{eq:Ls_local}
\end{equation}
where the zero-temperature value of the local inductance reads
$L_{i}(0) = L_s(0)\,\frac{T_{c, max}}{T_{c,i}}$, with $L_s(0)$ the reference inductance at the magic-angle value $T_{c, max}$. The rescaling of the local kinetic inductance ensures that bonds with higher $T_{c,i}$ carry a proportionally larger superfluid density.

For $T > T_{c,i}$, the bond is in the normal state and its 
resistance follows the Halperin-Nelson (HN) form~\cite{halperin1979resistive},
\begin{equation}
    R_i(T) = \frac{R_N}{1 + \xi_{\mathrm{\rm HN},i}(T)^2};
    \, 
    \xi_{\mathrm{\rm HN},i}(T) = \frac{2}{A_{\rm HN}}
    \sinh\!\left(\frac{b_{\mathrm{\rm HN}}}{\sqrt{t_i}}\right),
    \label{eq:HN_local}
\end{equation}
where $t_i = (T - T_{c,i})/T_{c,i}$ is the local 
reduced temperature, $A_{\rm HN}$ is a nonuniversal amplitude, and 
$b_{\mathrm{\rm HN}} \approx 2\alpha\sqrt{t_{c}}$ with 
$t_{c} = (T_{c,i}^0 - T_{c,i})/T_{c,i}$ the 
reduced separation between the local mean-field and BKT temperatures~\cite{benfatto2009broadening}. 
Since, as shown above, the zero-temperature inductance is rescaled inversely to the local critical temperature 
one can show that the ratio $T_{c,i}/T_{c,i}^0$ appearing in the reduced critical temperature is the same for every bond, i.e. $t_{c,i}\equiv t_c$. In the normal state, the superfluid stiffness vanishes and the inductive response reduces to a metallic residual inductance $L_i(T) = L_m$.

At each temperature, the network is solved numerically for the global impedance 
$Z_\mathrm{tot}(T,\omega_0) = R_\mathrm{tot} + i\omega_0 L_\mathrm{tot}$, 
from which we extract the DC resistivity, and the real and imaginary parts of
the complex optical conductivity,
\begin{equation}
    \sigma_{\rm tot}(T,\omega_0) = Z_\mathrm{tot}^{-1}(T,\omega_0) 
    = \sigma_1(T,\omega_0) - i\sigma_2(T,\omega_0)
    \label{eq:conductivity}
\end{equation}
encoding the superfluid stiffness via 
Eq.~\eqref{Eq:RelationSuperfluidStiffnesstoConductivity}. 
All observables are averaged over 100 disorder realizations, 
parametrized by the mean twist angle $\theta_0$ and its 
variance $\mathrm{Var}[\theta]$. We adopt $\omega_0 = 2\times10^9\,\mathrm{s}^{-1}$, 
$L_s(0) = 1\,\mathrm{nH}$, $L_m=10\,$nH, 
$R_N = 2\,\mathrm{k}\Omega$, $A_{\rm HN} = 1$, 
and $\alpha = 0.2$, so that $b_{HN}=0.15$, and compare our results directly with microwave resonance 
transport experiments~\cite{singh2018competition, Venditti2023RIN, Weitzel2023, 
Mallik2022, Banerjee2025}.

\begin{figure*}
    \centering
    \includegraphics[width=\linewidth]{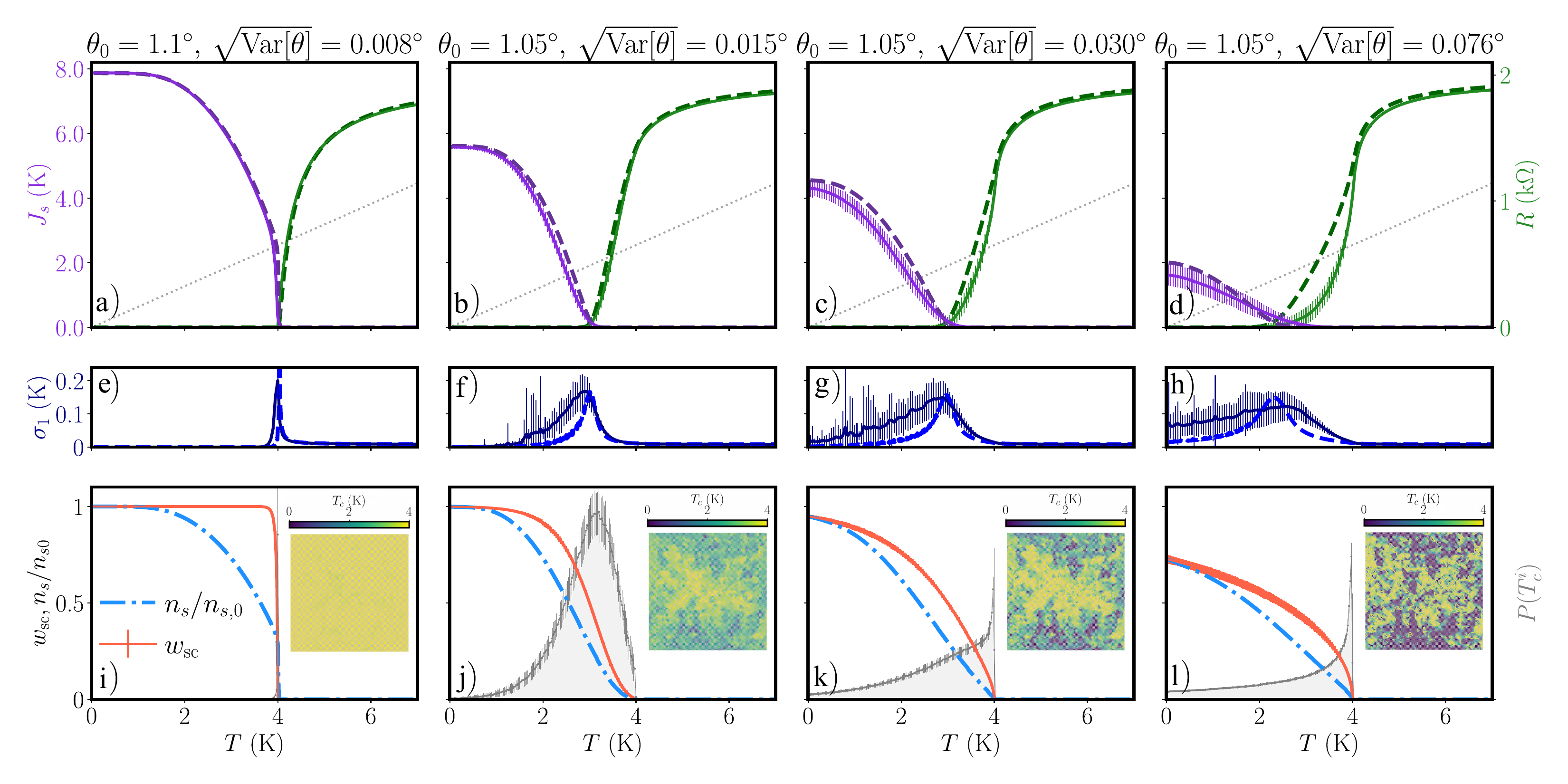}
    \caption{ 
    Results tuning the average $\theta_0$ angle and the width $\sqrt{{\rm Var}[\theta]}$ of the twist angle disorder.
    (a-d) DC resistivity (right axis, green) and superfluid stiffness (left axis, purple), within the exact RIN model (full lines) and in EMT (dashed).
    The dotted line is the BKT universal critical line $2T/\pi$.
    As disorder increases, the stiffness jump is progressively smeared, and $J_s$ crosses the $2T/\pi$ line continuously, with no jump visible. 
    Eventually, in the percolative regime, the crossing no longer coincides with the loss of global phase coherence, so it can no longer be used to define a meaningful $T_c$.
    (e-h) Optical conductivity $\sigma_1$ as a function of the temperature (RIN results in full lines, EMT in dashed). 
    As $\sqrt{{\rm Var}[\theta]}$ increases, the EMT fails in describing the broadening of the peak caused by long-range correlations.
    (i-l) Fraction of superconducting bonds $w_{\rm sc}$ (left axis, solid orange curve), superfluid density $n_s$ (left axis, dash-dotted light blue curve), and distributions of the local critical temperatures $P(T_c^i)$ (filled grey curves). Insets: maps of the critical temperatures. 
    $\theta_0=\theta_{\rm MA}$ and a very narrow $P(T_c^i)$ (i), results in a homogeneous $T_c$ map, which in turn gives a clear BKT signature in $J_s$ (a) and a sharp peak of $\sigma_1$ (e). 
    All error bars correspond to disorder configurations. 
    }
    \label{fig:RINpanel_vary_variance}
\end{figure*}

{\em RIN numerical results ---} 
Fig.~\ref{fig:RINpanel_vary_variance} presents 
our main results. 
The DC resistance, the superfluid stiffness, and the real part 
of the optical conductivity are shown as a function of temperature, for different values of 
the twist-angle variance at fixed mean twist angle $\theta_0 = 1.05^\circ$ 
and impurity concentration $n_\mathrm{imp} = 0.1$. 
We tune the width of the twist angle distribution up to $\sqrt{{\rm Var}[\theta]}=0.076^\circ$, comparable with the observations~\cite{Beechem2014,Uri2020,Turkel2022}. 
These are compared with the clean limit, shown in panels (a), (e), and (i) for $\theta_0=\theta_{\rm MA}=1.1^\circ$ and Var$[\theta]\approx 5 \cdot 10^{-5}$, where the canonical BKT picture is recovered. 
Here, the DC resistance is rounded by amplitude and phase fluctuations but vanishes sharply at the critical point. The superfluid stiffness follows the standard $s$-wave BCS temperature dependence at low temperatures and drops discontinuously to zero upon crossing the BKT critical line (dotted gray line). 
The real part of the optical conductivity (e) is sharply peaked at $T_c$ and vanishingly small both above and below it.

As $\theta_0 = 1.05^\circ$ and the twist-angle variance is increased, all three observables are progressively 
modified. Looking at panels (b)-(d), the DC resistance develops a low-temperature tail that broadens with increasing inhomogeneity. In the limit of strongly filamentary disorder, it can saturate to a finite value as a result of failed percolation of the
superconducting network~\cite{bucheli2013metal,dezi2018negative, Supplementary}. The discontinuous BKT jump in the superfluid 
stiffness smears into a smooth crossover, while its low-temperature profile is strongly renormalized: the characteristic exponential flatness of an $s$-wave condensate is progressively washed out, giving way to an approximately linear-in-$T$ suppression in panel (d) that closely resembles the behavior expected for a nodal superconductor. 
This is the first central result of our work: twist-angle inhomogeneity, naturally correlated in space, can reproduce the low-temperature stiffness signature of nodal pairing, irrespective of the actual gap symmetry.
The real part of the optical conductivity 
$\sigma_1(\omega_0,T)$, shown in panels (f-h), tells a complementary story: its peak broadens and 
acquires a finite low-temperature residual, reflecting the growing non-superconducting fraction of the network.
We stress that the results in Fig.\ref{fig:RINpanel_vary_variance} are averaged over 100 disorder realizations. Since however sample-to-sample fluctuations are sizable, a real device in a single disorder configuration could already display nodal-like features at smaller Var[$\theta$]~\cite{Supplementary}.

This last observation points to an important distinction. While in a clean BKT superconductor, the superfluid stiffness $J_s$ and the superfluid density $n_s$ are equivalent quantities, in the presence 
of disorder, they are not.
The global superfluid {\em stiffness} is in fact tied to the phase coherence of the condensate. This is equivalent to the helicity modulus, i.e., the second derivative of the free energy with respect to the twist of a phase, and thus to $\sigma_2\propto L_K^{-1}$, i.e., a direct measurement of the diamagnetic response of the superconductor.
A {\em density} instead, by definition, refers to a quantity per unit volume.
In panels (i)-(l) of Fig. ~\ref{fig:RINpanel_vary_variance}, we show the density of superconducting bonds $w_\mathrm{\rm sc}(T)=w_{\mathrm{sc},0} \int_T^\infty P(T_c)dT_c$ (in red), the spatially averaged local superfluid density $n_s(T)=\int_T  f_{\rm BCS}(T,T_{c,i}^0) P(T_c^i) dT_c^i$ (light blue), and the distributions of local critical temperatures $P(T_c)$ (gray shaded areas). 
Except for the clean limit, the resulting $n_s$ are clearly not proportional to $J_s$, which 
is sensitive to the spatial connectivity of the system. 
The nonequivalence between $J_s$ and $n_s$ is encoded in the broadening of $\sigma_1$, which makes it a practical experimental probe for the percolative nature of the transition. 
In the insets of panels (i)-(l) we show typical disorder configurations.

To disentangle the role of the disorder distribution from that of its spatial 
structure, we compare the full RIN results with estimates obtained within effective medium theory (EMT), see the End Matter. 
Since the EMT replaces the inhomogeneous network with a uniform 
medium, it captures the effect of the broadened $T_c$ distribution but discards all geometric information about spatial correlations. 
Wherever the exact and EMT solutions of the RIN agree, the response is governed by the distribution of local $T_c$'s alone; wherever they disagree, the excess is a direct fingerprint of strong spatial correlations leading to filamentary connectivity. 
The EMT results are shown as dashed lines in Fig.~\ref{fig:RINpanel_vary_variance}. 
Notice how the EMT results spectacularly fail in the description of $\sigma_1$, making this observable a smoking gun for long-range correlated disorder. That is the second key result of this work.

The robustness of these results  
are discussed in the Supplemental Material~\cite{Supplementary}.

{\em Conclusions and outlook ---} 
{In this Letter, we have shown that twist-angle inhomogeneities in moiré superconductors are not incidental: they are an inevitable consequence of the elastic sensitivity of the lattice to atomic-scale defects, and they are logarithmically long-range correlated.
Mapping these inhomogeneities onto a landscape of local BKT transition temperatures and modeling the system as a random impedance network, we have demonstrated that correlated disorder has three distinct and experimentally consequential effects. 

First, it broadens the BKT transition into a smooth crossover, eventually driving the 
system toward a percolative transition among superconducting islands. 
In this regime, the point where $2T/\pi$ crosses the superfluid stiffness $J_s$ can no longer be identified with the critical point for global phase coherence.

Second, 
and more subtly, correlated disorder produces a power-law suppression of the 
superfluid stiffness at low temperatures that closely mimics the signature of 
nodal superconductivity, regardless of the actual gap symmetry.

Finally, our analysis further identifies the real part of the optical conductivity $\sigma_1(\omega_0, T)$ 
as a key diagnostic observable that has been largely overlooked, especially in the moiré 
context. In a clean superconductor, $\sigma_1$ displays a sharp peak at 
$T_c$ and is negligibly small elsewhere. The broadening of this peak, and particularly its finite residual value at low temperatures, directly quantifies 
the level of spatially-correlated inhomogeneity. If $\sigma_1$ 
exhibits only a sharp peak at $T_c$, the sample can be considered 
sufficiently homogeneous.

Our theoretical framework can help shed light on the nature of the superconducting order parameter in graphene multilayers.
Resonant microwave experiments reported nodal-like features of the stiffness~\cite{Tanaka2025,Banerjee2025}; however, neither study showed the corresponding $\sigma_1$, necessary to draw firm conclusions about the exotic/nodal nature of the superconducting state.
In contrast, recent observations of $2e$-periodic conductance in tBG superconducting islands would suggest a fully gapped state~\cite{Mestre-Tor2026}. 
Furthermore, our picture is consistent with the observation of the superconducting Josephson diode effect in magic-angle tBG~\cite{Rothstein2026}, where the nonreciprocal supercurrent was traced to a non-uniform supercurrent distribution induced by twist-angle inhomogeneity. 
More generally, our analysis gives a broad understanding of the critical temperature $T_c$ and its dependence on the distribution of local twist angles, providing a more quantitative explanation of the reproducibility issues reported in tBG devices~\cite{lau_reproducibility_2022}, in particular why nominally identical samples may or may not exhibit superconductivity. 
At the same time, the elastic origin of these correlations points to a unifying description of long-range disorder in other 2D systems, as recently shown also for electronic nematic order~\cite{Meese_compatible_2026, Meese_theorynematic_2026}.

The natural step forward is to move beyond spatially averaged transport and access 
the local electromagnetic response directly. Local probes can provide complementary insight into both the order-parameter symmetry and the underlying disorder landscape. 
Spectroscopically, scanning tunneling microscopy (STM) maps the local gap: in twisted trilayer graphene, it revealed U- and V-shaped superconducting gaps as a function 
of filling, together with an emergent pseudogap phase~\cite{Kim2022}, and in tBG a putative unconventional superconductor with a pseudogap~\cite{Oh2021}. 
Interestingly, such a pseudogap can naturally arise in a spatially inhomogeneous landscape where locally superconducting islands retain a well-defined 
pairing gap even where global phase coherence is suppressed, so that a local probe 
registers a spectral gap in the absence of coherent bulk superconductivity~\cite{bucheli2015pseudo}. 
Scanning microwave 
impedance microscopy~\cite{Barber2022} and nano-SQUID 
magnetometry~\cite{Uri2020,Jha2025} are well suited to resolve the nanoscale 
inhomogeneity landscape of moiré systems, and simultaneous local and global 
measurements would allow one to correlate the twist-angle map with the 
superconducting response of the same sample. Such measurements would also 
allow for a direct distinction between the superfluid stiffness $J_s$ and the 
superfluid density $n_s$.
The combination of local and transport probes could also be relevant to investigating the quantum-geometrical contribution to the (local) kinetic inductance \cite{Trm2022, Tian2023, Jha2025}. Finally, local shot noise spectroscopy \cite{Blanter.2000} can elucidate the difference between the superconducting and normal domains. This wide variety of local probes would allow future researchers to clearly distinguish disorder effects from intrinsic properties of superconductors -- including the order parameter symmetry.

\vspace{0.25cm}
{\em Acknowledgments---}We thank Sergio Caprara, Marco Grilli, Dima Efetov, Johannes Knolle, Kaveh Lahabi, Margherita Melegari, Milan Allan, and Rafael Fernandes, for fruitful discussions. 
G.V. acknowledges the Swiss National Science Foundation (SNSF) via Swiss Postdoctoral Fellowship Grant No.\ TMPFP2~224637. 
L.R. acknowledges the SNSF via Starting Grant TMSGI2~211296.
I.M. acknowledges financial support by the SNSF via the Swiss Postdoctoral Fellowship Grant No. TMPFP2~217204.


\bibliography{biblio}

\onecolumngrid

\vspace{1em}
\begin{center}
\textbf{End Matter}
\end{center}
\vspace{1em}

\twocolumngrid

\emph{Effective medium approximation ---}
The RIN model can be solved in the EMT approximation \cite{venditti2020superfluid, venditti2021finite}, implicitly assuming uncorrelated inhomogeneities. 
It is convenient to split the bonds into two species: a fraction $w_{\mathrm{sc},0}$, that
eventually becomes superconducting, and the residual metallic matrix, of fraction $w_m=1-w_{\mathrm{sc},0}$. While the latter stays metallic at all temperatures with impedance $z_m=R_N+i\omega L_m$, the impedance of former depends on the local BKT temperatures, which are distributed according to the same $P(T_c)$ induced from the twist-angle distributions $P(\theta)$, and carefully renormalized $\int_0^\infty P(T_c)\,dT_c=1$.
Each bond species is an effective-medium component, so the EMT equation reads
\begin{widetext}
\begin{equation}
  w_{\mathrm{sc},0}\!\int_0^\infty\!dT_c\,P(T_c)\,
  \frac{z_{em}-z_s(\omega,T;T_c)}
       {z_{em}+z_s(\omega,T;T_c)}
  +(1-w_{\mathrm{sc},0})\,\frac{z_{em}-z_m}{z_{em}+z_m}=0. 
  \label{eq:twoclass}
\end{equation}
\end{widetext}
where the local impedances of the SC bonds read: 
\begin{equation}
  z(\omega,T;T_c)=
  \begin{cases}
    z_s=i\omega L_s(T),         & T<T_c,\\[2pt]
    z_m=R_N(T)+i\omega L_m,      & T>T_c.
  \end{cases}
\end{equation}
The temperature dependence of the local inductance and resistance are the same as defined in the main text:
\begin{equation}
    L_s(T) = \frac{\hbar^2}{4e^2 k_B J_{s}(T)},
    \label{eq:Ls_localEMT}
\end{equation}
where:
\begin{align}
J_s(T;T_c)=J_s(0) g(\tau_i)
    \tanh\!\left(\frac{1.764\,g(\tau_i)}{2\tau_i}\right),
  \label{eq:ourLs}
\end{align}
$J_s(0)=\frac{\hbar^2 }{4e^2k_B L_s(0)}\frac{T_c}{T_{c, max}}$ and $g(\tau)=\sqrt{\cos(\pi\tau^2/2)}$.  

\begin{equation}
    R_N(T;T_c)=\frac{R_N}{1+\xi_{\mathrm{HN}}^{2}},
  \qquad
  \xi_{\mathrm{HN}}=\frac{2}{A_{\mathrm{HN}}}\sinh\!\frac{b_{\mathrm{HN}}}{\sqrt{t}},
  \label{eq:ourR}
\end{equation}
with $t=\frac{T-T_c}{T_c}$.

At each temperature, Eq.~\eqref{eq:twoclass} is solved numerically for the complex effective impedance $z_{em}(\omega,T)$, from which the DC resistance, the real part of the optical conductivity $\sigma_1$, and the superfluid stiffness are extracted.

Because the EMT mixes the two bond species at random, it is accurate whenever the disorder is spatially homogeneous --- that is, when it does not organize into the extended, filamentary structures that long-range correlations generate. 
In 2D, the percolating threshold of the EMT is $w_{\rm sp}=0.5$, i.e., for the SC transition to occur the system requires at least half of the bonds to become SC.
The exact RIN solution can instead sustain a transition even for $w_{\rm sc}<w_{\rm sp}$, as soon as a percolating path connects the two edges of the sample; the long, filamentary clusters that carry this path are also what generate the pronounced low-$T$ tails in the DC resistivity, which the EMT cannot reproduce. 
At the same time, in samples with $w_{\rm sc}<w_{\rm sp}$, the EMT underestimates $J_s$ and its onset temperature.
In the Supplemental Material \cite{Supplementary}, we present the EMT result for a wide range of parameters $\theta_0$, and $\sqrt{{\rm Var}[\theta]}$.

\emph{Gapped vs nodal order parameter --- }
\begin{figure}
    \centering
    \includegraphics[width=\linewidth]{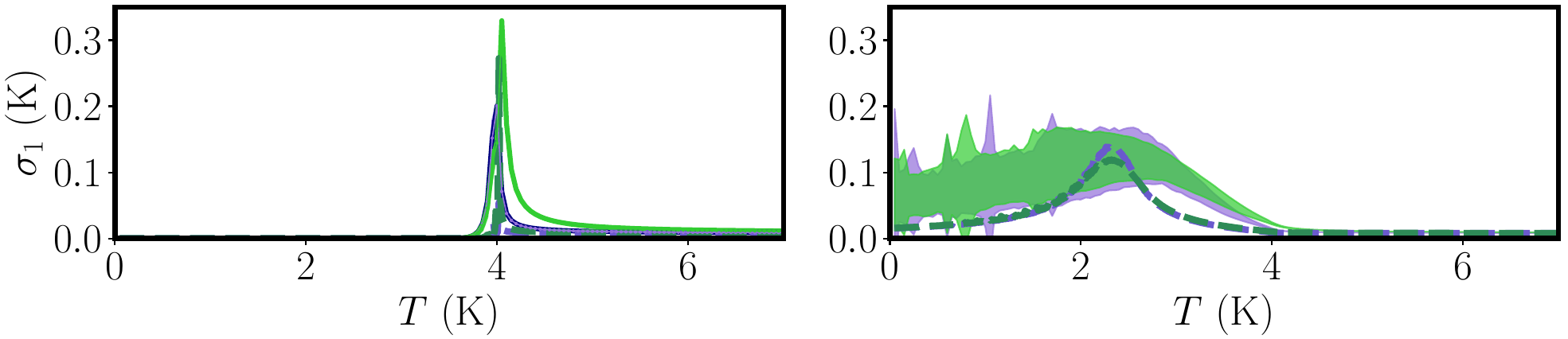}
    \caption{Comparison of $\sigma_1$ of the computed exact RIN solution (filled lines) and EMT approximation (dashed lines) between gapped (violet) and nodal (green) profiles of the superconducting order parameter. We compare two possible twist angle distributions. 
    The width of the exact RIN solutions represents the error bars given by the disorder realizations. }
    \label{fig:BCS-vs-nodal_s1}
\end{figure}
In Fig.~\ref{fig:BCS-vs-nodal}, we showed the superfluid stiffness in two paradigmatic twist angle realizations: a very clean system, i.e., $\theta_0=1.1^\circ$ and $\sqrt{{\rm Var}[\theta]}=0.008^\circ$, and a more realistic sample with $\theta_0=1.05^\circ$ and $\sqrt{{\rm Var}[\theta]}=0.076^\circ$.
For the sake of completeness, we report here in Fig.~\ref{fig:BCS-vs-nodal_s1} the comparison of the corresponding $\sigma_1$.
The gapped profile $f_{\rm BCS}(T,T_{c,i}^0)$ is again the one in Eq.~\eqref{eq:Js_local} and \eqref{eq:ourLs};
the nodal profile considered here is instead simply
\begin{equation}
    f_{\rm nodal}(T,T_{c,i}^0)=1-\frac{T}{T_{c,i}^0}. 
\end{equation}
In the clean case, $\sigma_1$ shows the paradigmatic peak in $T_c$, dropping to zero at lower temperatures. This correspond to a clear distinction between the gapped and the nodal $J_s$ profiles.
In the inhomogeneous case, the EMT fails. At the same time the broadening of $\sigma_1$ and its substantial low-$T$ residual value mirror the absence of a clear distinction between the gapped and the nodal order parameters.

\clearpage
\onecolumngrid
\setcounter{equation}{0}
\setcounter{figure}{0}
\setcounter{table}{0}
\setcounter{section}{0}
\renewcommand{\theequation}{S\arabic{equation}}
\renewcommand{\thefigure}{S\arabic{figure}}
\renewcommand{\thetable}{S\arabic{table}}
\renewcommand{\thesection}{S\Roman{section}}
\SMbibstart

\begin{center}
{\large\bfseries Supplemental Material:\\[2pt] How Twist Angle Inhomogeneity Masks the Superconducting Order Parameter Symmetry}\\[1em]
Ilaria Maccari, Louk Rademaker, and Giulia Venditti
\end{center}
\vspace{1em}
\section{Elastic theory of solids}

Here we provide some more details on the derivation of the logarithmic correlations in the twist angle inhomogeneity. Following Landau~\citeSM{landau1986theory}, the condition for equilibrium for an elastic solid reads
\begin{equation}
	{\bf F}_i \equiv \frac{\partial \sigma_{ik}}{\partial x_k} =0,
\end{equation}
where $\sigma_{ik}$ is the {\em stress tensor}. The expression ${\bf F}_i$ is the force per unit volume. The stress tensor for a general body can be expressed in terms of the strain tensor $u_{ik}$ as
\begin{eqnarray}
	\sigma_{ik} = \frac{E}{1+\nu} \left( u_{ik} + \frac{\nu}{1 - 2\nu} u_{ll} \delta_{ik} \right)
\end{eqnarray}
where $\nu$ is Poisson's ratio and $E$ is Young's modulus. For small deformations, the strain tensor is related to the displacements $\vec{u}$ through
\begin{equation}
	u_{ik} = \frac{1}{2} \left( \frac{\partial u_i}{\partial x_k}+ \frac{\partial u_k}{\partial x_i} \right).
\end{equation}
Combining all these equations gives the equilibrium condition for an elastic body,
\begin{equation}
	\Delta \vec{u} + \frac{1}{1 - 2 \nu} \vec{\nabla} (\vec{\nabla} \cdot \vec{u} ) = 0.
	\label{SM-Eq:EquilibriumDisplacement}
\end{equation}
We next imagine there are random forces $\vec{f}$ acting on the solid medium -- for example due to impurities in the substrate -- such that the condition for elastic equilibrium becomes
\begin{equation}
	\frac{\partial \sigma_{ik}}{\partial x_k} + \rho \vec{f} = 0
\end{equation}
where $\rho$ is the density of the body. Equation Eq.~\eqref{SM-Eq:EquilibriumDisplacement} now becomes
\begin{equation}
	\Delta \vec{u} + \frac{1}{1 - 2 \nu} \vec{\nabla} (\vec{\nabla} \cdot \vec{u} ) = -\rho \vec{f} \frac{2 (1 + \nu)}{E}
\label{SM-Eq:ForcesDisplacement}
\end{equation}
We can solve this using the Green's function method. For simplicity we write $\vec{F} =  \rho \vec{f} \frac{2 (1 + \nu)}{E}$. Then Fourier transforming $\vec{u}(\vec{x}) = \int \frac{d^2k}{(2\pi)^2} \vec{u}(\vec{k}) e^{i \vec{k} \cdot \vec{x}}$ yields, for a 2d material,
\begin{equation}
	\begin{pmatrix}
		k^2 + \frac{1}{1 - 2 \nu} k_x^2 & \frac{1}{1 - 2\nu} k_x k_y \\
		\frac{1}{1 - 2\nu} k_x k_y & k^2 + \frac{1}{1 - 2 \nu} k_y^2
	\end{pmatrix}
	\begin{pmatrix}
		u_x (\vec{k}) \\
		u_y (\vec{k})
	\end{pmatrix}
	= 
	\begin{pmatrix}
		F_{x} (\vec{k}) \\ F_{y} (\vec{k})
	\end{pmatrix}.
\end{equation}
Inverting this matrix yields
\begin{equation}
	\begin{pmatrix}
		u_x (\vec{k}) \\
		u_y (\vec{k})
	\end{pmatrix}
	= 
	\frac{1}{2 k^4 (1-\nu )}
	\begin{pmatrix}
		k_x^2 (1- 2\nu) + 2 k_y^2 (1-\nu) & {-}k_x k_y \\
		 {-}k_xk_y & k_y^2 (1- 2\nu) + 2 k_x^2 (1-\nu)
	\end{pmatrix}
	\begin{pmatrix}
		F_{x} (\vec{k}) \\ F_{y} (\vec{k})
	\end{pmatrix}.	
\end{equation}
The above equation has the shape $\vec{u} (\vec{k}) = \hat{G}(\vec{k}) \vec{F}(\vec{k})$, with
\begin{align}
	\hat{G}(\vec{k}) &=
		\frac{1}{2 k^4 (1-\nu )}
	\begin{pmatrix}
		k_x^2 (1- 2\nu) + 2 k_y^2 (1-\nu) & {-}k_x k_y \\
		 {-}k_xk_y & k_y^2 (1- 2\nu) + 2 k_x^2 (1-\nu)
	\end{pmatrix}.
	\label{SM-Eq:GreensFunctionMomentumSpace}
\end{align}
This can be Fourier transformed back to real space to yield a convolution
\begin{equation}
	\vec{u}(\vec{x}) = \int d^2x' \hat{G}(\vec{x} - \vec{x'}) \vec{F}(\vec{x}').
\end{equation}
Explicitly performing the inverse Fourier transform yields
\begin{eqnarray}
	G_{xx} (\vec{x}) & = & 
		\int \frac{d^2k}{(2\pi)^2} 
			\frac{k_x^2 (1- 2\nu) + 2 k_y^2 (1-\nu)}{2 k^4 (1-\nu )} e^{i \vec{k} \vec{x}} \\
		&=&- \frac{3-4\nu}{16 \pi(1-\nu)} \left( 2 \gamma + \log |\vec{x}|^2 \right)
		+ \frac{x^2}{8 \pi |\vec{x}|^2 (1 - \nu)}
		\\
	G_{xy} (\vec{x}) & = & 
		\int \frac{d^2k}{(2\pi)^2} \frac{-k_xk_y}{2 k^4 (1-\nu )} e^{i \vec{k} \vec{x}} \\
		& = & \frac{xy}{8 \pi |\vec{x}|^2 (1 - \nu)}
\end{eqnarray}
where $\gamma \approx 0.577216$ is Euler's constant. When considering a finite size with open or periodic boundary conditions, the Fourier transform in the above equation should restrict the relevant momenta.

Next, we analyze how the elasticity of the solid leads to correlations in the twist angle inhomogeneity. Following the continuum theory of twisted layers -- see Eq. (32) of Ref.~\citeSM{Balents2019} -- we know that a rigidly twisted system has displacement
\begin{equation}
	\vec{u} (\vec{x}) = \frac{1}{2} \delta\theta \hat{z} \times \vec{x} = \frac{\delta\theta}{2}  (-y, x,0)
\end{equation}
A rigid twist does not have any divergence; it is purely a curl. We invert the above relation to define the local twist angle as
\begin{equation}
\delta\theta(\vec{x}) \equiv \hat{z} \cdot ( \vec{\nabla} \times \vec{u}(\vec{x}) )
\label{SM-Eq:ThetaField}
\end{equation}
This allows us now to calculate the correlations of the twist angle inhomogeneity,
\begin{equation}
	C_\theta (\vec{x}_1 - \vec{x}_2) = \langle \delta\theta(\vec{x}_1) \delta\theta(\vec{x}_2) \rangle.
\end{equation}
Using the relations above, we can write that given a certain force distribution $\vec{F}(\vec{R})$, the local twist angle is
\begin{align}
	\delta\theta (\vec{x}) & =  \sum_{\vec{k} \vec{R}} \epsilon_{ab} \nabla_a G_{bc}(\vec{k}) F_c (\vec{R}) 
		e^{-i \vec{k} \cdot (\vec{x} - \vec{R})} \\
		&=  -i \sum_{\vec{k} \vec{R}} \epsilon_{ab} k_a G_{bc}(\vec{k}) F_c (\vec{R}) 
		e^{-i \vec{k} \cdot (\vec{x} - \vec{R})}
\end{align}
where $\epsilon_{ab}$ is the antisymmetric tensor. The correlation function, obtained by disorder averaging, reads
\begin{align}
	C_\theta (\vec{x}_1 - \vec{x}_2) &= -
		\sum_{\vec{k} \vec{R}\vec{k}' \vec{R}'} \langle \epsilon_{ab} k_a G_{bc}(\vec{k}) F_c (\vec{R}) 
		\epsilon_{de} k'_d G_{ef}(\vec{k}') F_f (\vec{R}') 
		e^{-i \vec{k} \cdot (\vec{x}_1 - \vec{R})}
		e^{-i \vec{k}' \cdot (\vec{x}_2 - \vec{R}')} \rangle.
\end{align}
The forces are assumed to be random and uncorrelated,
\begin{align}
	\langle F_c (\vec{R})  F_f (\vec{R}') \rangle \propto \delta_{cf} \delta (\vec{R} - \vec{R}')
\end{align}
so that the correlation function becomes
\begin{align}
	C_\theta  (\vec{x}_1 - \vec{x}_2) &\propto  -
		\sum_{\vec{k} \vec{R}\vec{k}' } \epsilon_{ab} k_a G_{bc}(\vec{k}) 
		\epsilon_{de} k'_d G_{ec}(\vec{k}') 
		\; e^{i \vec{R} \cdot (\vec{k} + \vec{k'})}
		e^{-i \vec{k} \cdot \vec{x}_1}
		e^{-i \vec{k}' \cdot \vec{x}_2}
\end{align}
Integrating over $\vec{R}$ yields $\vec{k}' = - \vec{k}$, so that we get
\begin{align}
	C_\theta  (\vec{x}_1 - \vec{x}_2) &\propto
		 \sum_{\vec{k}} \epsilon_{ab} k_a G_{bc}(\vec{k}) 
		\epsilon_{de} k'_d G_{ec}(-\vec{k}) 
		\; \; e^{-i \vec{k} \cdot (\vec{x}_1 - \vec{x}_2)} \\
	&= \sum_{\vec{k}} \vec{k} \cdot \hat{\epsilon}  \cdot \hat{G}_{\vec k}  \cdot \hat{G}^\intercal_{-\vec{k}}  \cdot \hat{\epsilon}^{\intercal} \cdot \vec{k}
	\; \; e^{-i \vec{k} \cdot (\vec{x}_1 - \vec{x}_2)} 
\end{align}
In the last line, we wrote the term inside the inverse Fourier transform as a matrix product. We can use the Green's function of Eq.~\eqref{SM-Eq:GreensFunctionMomentumSpace} to obtain
\begin{align}
	C_\theta  (\vec{x}_1 - \vec{x}_2) &\propto  \sum_{\vec{k}} \frac{1}{|\vec{k}|^2} \; e^{-i \vec{k} \cdot (\vec{x}_1 - \vec{x}_2)} 
\end{align}
The surprisingly simple result is that the Fourier transform of the twist angle correlation function is $C_\theta(k) \propto 1/k^2$. The inverse Fourier transform can be done by noticing that the twist angle correlation function should be a solution to the 2d Laplace equation, which yields
\begin{align}
	C_\theta (\vec{r}) \propto  {\rm constant} - \log |\vec{r}|.
\end{align}

\section{Analysis of the numerical $\delta\theta$ maps}

We compute the $\delta\theta$ maps numerically by translating Eq.~\eqref{SM-Eq:ForcesDisplacement} in its discrete form on a square lattice with open boundary conditions.
For each realization of disorder, we randomly pick a fraction $n_{\rm imp}$ of sites where the local random forces ${\bf F}_i=(F_{x,i}, F_{y,i})$ act, with a random orientation in the plane, and a random strength uniformly distributed in $(0,F_{\rm max}]$, see Fig.~\ref{SM-fig:u-F-fields}a. The corresponding computed $\bf u$ fields and $\delta\theta$ maps are shown in Fig.~\ref{SM-fig:u-F-fields}b,c.
\begin{figure}
    \centering
        \includegraphics[width=\linewidth]{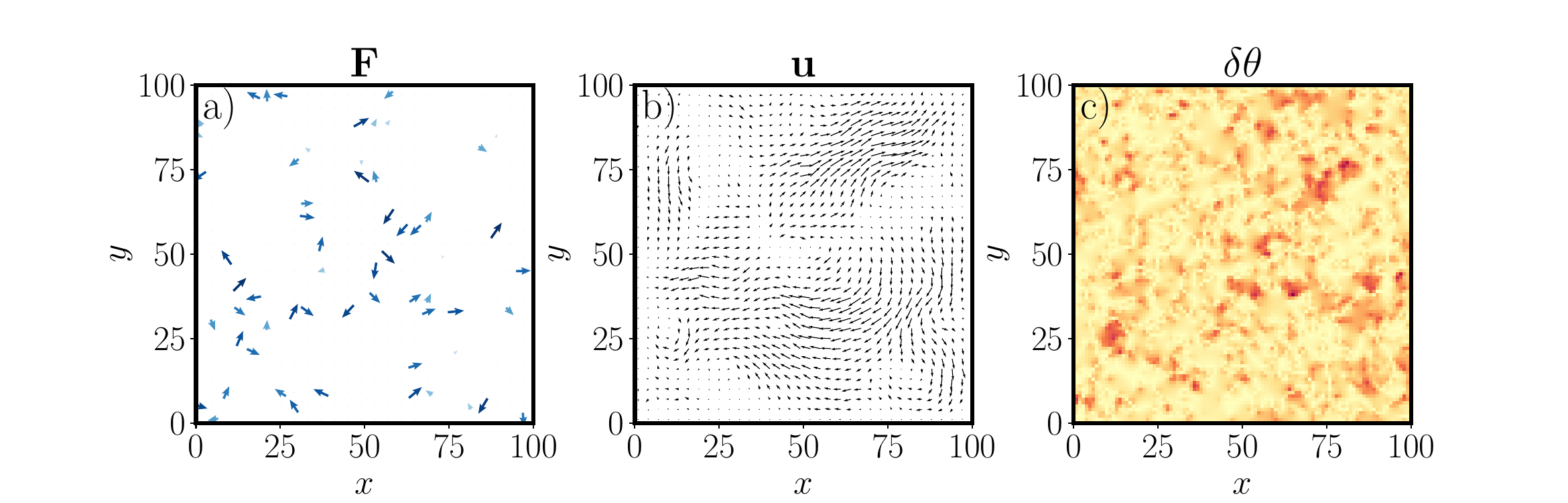}
    \caption{
    (a) Random coarse-grained local forces $\bf F$; their corresponding (b) $\bf u$ displacement field and (c) $\delta\theta$ maps, computed from the discretized elasticity equations in Eq.~\eqref{SM-Eq:ForcesDisplacement}. 
    We take $n_{\rm imp}=0.1$ on a $100\times 100$ square lattice. The local random forces have $F_{\rm max}=1$.}
    \label{SM-fig:u-F-fields}
\end{figure}

We show in Fig.~\ref{SM-fig:combined_corrfunc} the results of our numerical analysis on the twist angle distributions $P(\delta\theta)$ and correlation functions obtained at various $n_{\rm imp}$, keeping $F_{\rm max}$ fixed to 1 for simplicity.
The numerical correlation functions are fitted to $C_\theta(r)=-a\ln(r)+b$.
In the table in Fig.~\ref{SM-fig:combined_corrfunc}a, we present the obtained values of the numerical variance of the coarse-grained forces Var[$F$], as well as the log-fit slope $a$, the variance of the twist angle Var[$\theta$], and, for the sake of completeness, the Kurtosis $\mathcal{K}=\langle\delta\theta^4\rangle/\langle\delta\theta^2\rangle^2$ of the twist angle distributions $P(\delta\theta)$.
This former value indicates that $P(\delta\theta)$ crosses from a Laplacian-like form at $n_{\rm imp}=0.01$ where $\mathcal{K}\approx 6$, to a Gaussian at $n_{\rm imp}=0.01$, $\mathcal{K}\approx 3$. 
In light of the RIN calculations, once the variance Var[$\theta$] is fixed, the exact shape of the probability distribution won't affect the calculations (see below, Fig.~\ref{SM-fig:transport-vs-nimp}).
Indeed, Var($\theta$) scales linearly with $n_{\rm imp}$Var($F$) (Fig.~\ref{SM-fig:combined_corrfunc}b), making it one of the most relevant parameters at play.
Similarly, the logarithmic long-range correlation is not affected by $n_{\rm imp}$ alone, and it remains a constant (Fig.~\ref{SM-fig:combined_corrfunc}c).

\begin{figure}
\centering
\includegraphics[width=0.9\linewidth]{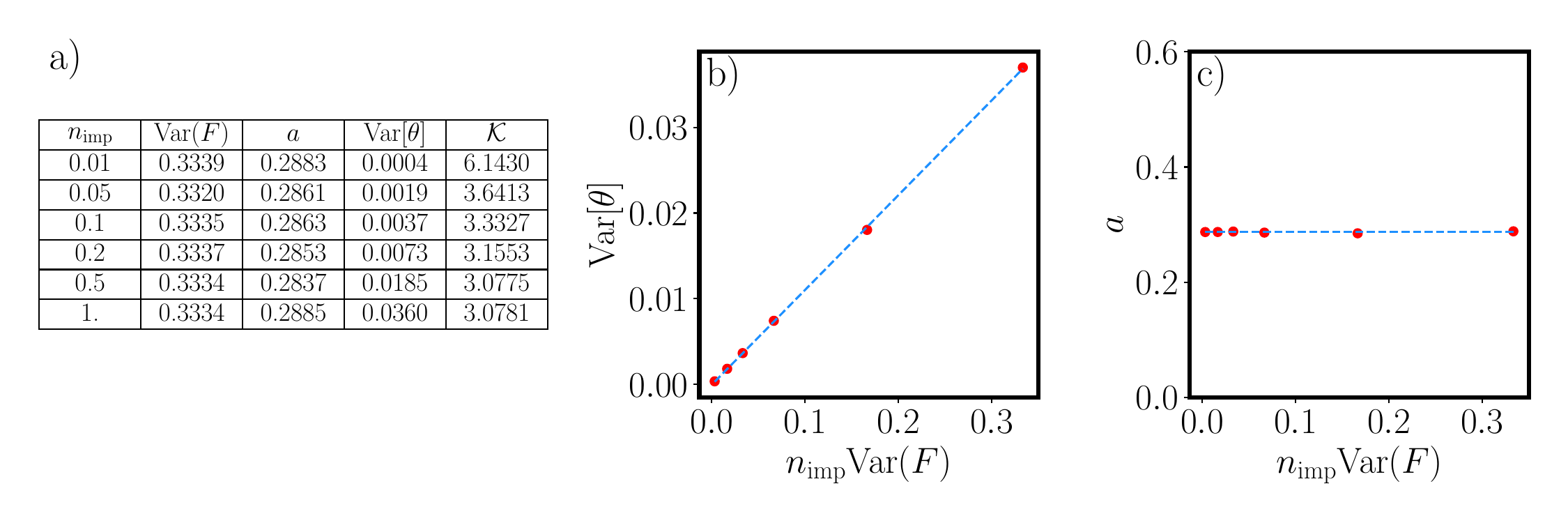}
\caption{(a) Numerical analysis of the correlation function $C_\theta({\bf r})=-a\ln |{\bf r}|+b$ for various $n_{\rm imp}$ and fixed local forces $|{\bf F}|$. The Kurtosis $\mathcal{K}$ indicates that the $P(\delta\theta)$ cross from a Laplacian-like form at $n_{\rm imp}=0.01$ ($\mathcal{K}\approx 6$), to a Gaussian at $n_{\rm imp}=1$ ($\mathcal{K}\approx 3$). (b) Var[$\theta$] scales linearly with $n_{\rm imp}$Var($F$). (c) The log-fit slope of $C_\theta(r)=-a\ln(r)+b$ remains constant.}
\label{SM-fig:combined_corrfunc}
\end{figure}

\section{Mapping $\delta\theta$ into $T_{c}$}

Once the  $\delta\theta$ maps are generated, we tune the width of $\delta\theta$ by a factor $r_{\rm maps}$ such that Var[$\theta$] is comparable with the observed variances of twist angles~\citeSM{Beechem2014,Uri2020,Turkel2022,Hu.2024.CPL} and we consider different average angles $\theta_0$.
We plot their distribution in Fig.~\ref{SM-fig:PTc}(a-d).

The $\delta \theta$ maps generated by the elasticity equations are defined on the sites of the lattice, whereas the RIN model is defined on the bonds. 
Hence, the critical temperature
\begin{equation}
    T_{c,i} = T_{c, {\rm max}} 
    \left( 1 - \frac{(\theta_i - \theta_{\rm MA})^2}{\Delta \theta^2}\right),
    \label{SM-eq:Tc_phenomen}
\end{equation}
with $\theta_i=\theta_0+r_{\rm maps}\delta\theta_i$ and $T_{c,\rm max}$, is associated with the two bonds along $x,y$ directions.
The induced probabilities on the critical temperatures $P(T_c)$ are computed numerically and plotted in Fig.~\ref{SM-fig:PTc}(e-h).

\begin{figure}[h!]
    \centering
    \includegraphics[width=\linewidth]{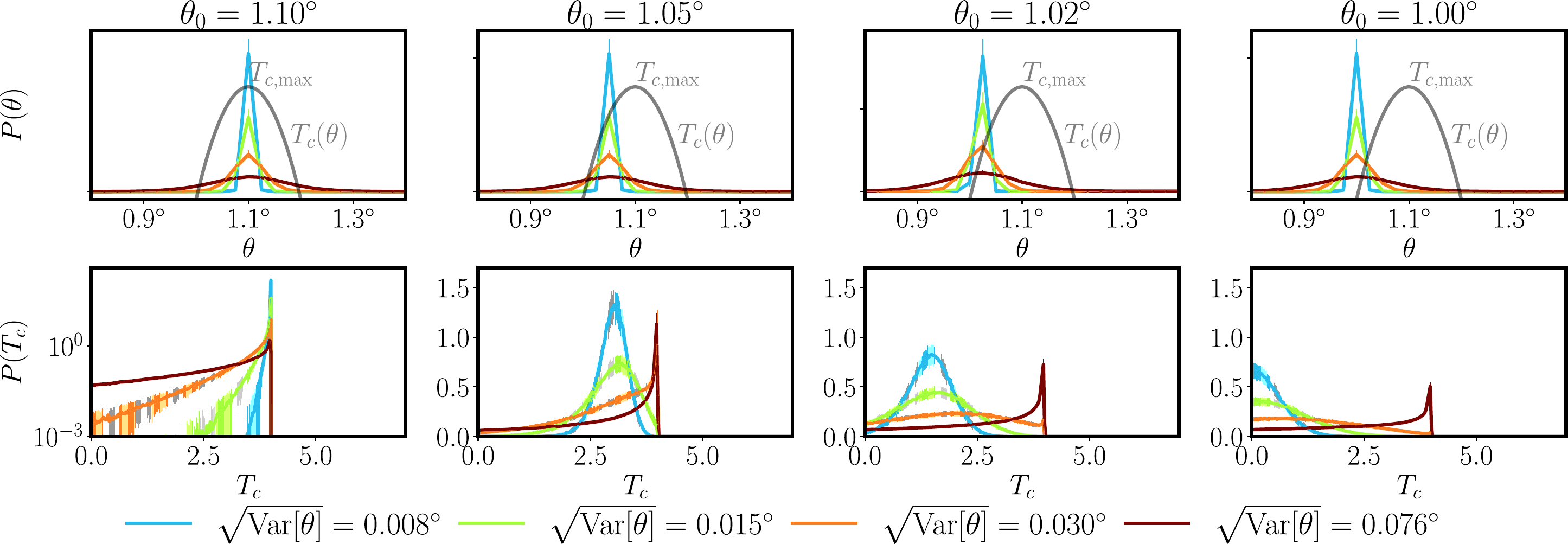}
    \caption{ 
    Upper panels: distributions of local twist angles $P(\theta)$ at various $\theta_0$ and $\sqrt{{\rm Var}[\theta]}$. 
    The grey dome is the phenomenological formula in Eq.~\eqref{SM-eq:Tc_phenomen}.
    Lower panels: induced $P(T_c)$ distributions.
    }
    \label{SM-fig:PTc}
\end{figure}

\section{Experimental $T_c(\theta)$ data from the literature}
\label{SM-sec:Tc-theta-literature}

To put the phenomenological form of Eq.~\eqref{SM-eq:Tc_phenomen} in context, we collect in
Table~\ref{SM-tab:Tc_theta_literature} the superconducting critical temperatures $T_c$ reported in the
magic-angle twisted bilayer graphene (MATBG) literature as a function of twist angle $\theta$, together
with the associated experimental uncertainties where these are reported. The data points were mentioned in the review of Ref.~\citeSM{Koblischka.2024}, and the original sources are Refs.~\citeSM{Cao2018,Saito2020,Lu2019,Stepanov2020}. Error bars on
$\theta$ and $T_c$ are shown only where explicitly reported.
Figure~\ref{SM-fig:Tc_theta_literature} plots this compilation together with the curve of
Eq.~\eqref{SM-eq:Tc_phenomen} for $T_{c,\rm max}=4.0~\mathrm{K}$, $\theta_{\rm MA}=1.1^\circ$, and
$\Delta\theta=0.1^\circ$.

\begin{table}[h]
\centering
\caption{Superconducting critical temperature $T_c$ versus twist angle $\theta$ reported in the MATBG
literature, as compiled in Ref.~\citeSM{Koblischka.2024} and cross-checked against the original
publications. Uncertainties are quoted only where explicitly reported by the source.}
\label{SM-tab:Tc_theta_literature}
\begin{tabular}{ccl}
\toprule
Twist angle $\theta$ ($^\circ$) & $T_c$ (K) & Reference \\
\midrule
$1.16$                 & $0.5$                          & Cao \emph{et al.} \citeSM{Cao2018} \\
$1.05$                 & $1.7$                          & Cao \emph{et al.} \citeSM{Cao2018} \\
$1.08 \pm 0.02$        & $2.3\; ^{+1.4}_{-0.4}$         & Saito \emph{et al.}, device 1 \citeSM{Saito2020} \\
$1.09 \pm 0.02$        & $2.4\; ^{+1.9}_{-1.6}$         & Saito \emph{et al.}, device 2 \citeSM{Saito2020} \\
$1.04 \pm 0.02$        & $1.3 \pm 0.6$                  & Saito \emph{et al.}, device 3 \citeSM{Saito2020} \\
$1.12 \pm 0.01$        & $4.0\; ^{+2.05}_{-2.4}$        & Saito \emph{et al.}, device 5 \citeSM{Saito2020} \\
$1.18 \pm 0.03$        & $0.6\; ^{+0.4}_{-0.45}$        & Saito \emph{et al.}, device 4 \citeSM{Saito2020} \\
$1.10$                 & $2.3$                          & Lu \emph{et al.}, max.\ $T_c$ \citeSM{Lu2019} \\
$1.15$                 & $0.9$                          & Stepanov \emph{et al.}, D1 \citeSM{Stepanov2020} \\
$1.04$                 & $0.4$                          & Stepanov \emph{et al.}, D2 \citeSM{Stepanov2020} \\
\bottomrule
\end{tabular}
\end{table}

\begin{figure}[h]
    \centering
    \includegraphics[width=0.5\linewidth]{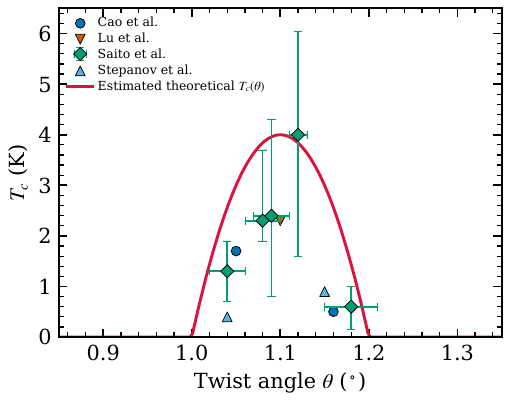}
    \caption{
    Superconducting critical temperature $T_c$ versus twist angle $\theta$ compiled from the MATBG
    literature (Table~\ref{SM-tab:Tc_theta_literature}), with error bars shown where reported by the source
    (Ref.~\citeSM{Saito2020}). The solid curve is the phenomenological form of Eq.~\eqref{SM-eq:Tc_phenomen},
    $T_c(\theta)=T_{c,\rm max}\left(1-(\theta-\theta_{\rm MA})^2/\Delta\theta^2\right)$, with
    $T_{c,\rm max}=4.0$~K, $\theta_{\rm MA}=1.1^\circ$, and $\Delta\theta=0.1^\circ$.
    }
    \label{SM-fig:Tc_theta_literature}
\end{figure}

\section{Random impedance network}

\subsection{Exact solution}

Kirchhoff and Ohm equations are implemented for a square lattice with open boundary conditions, where an external applied voltage $V_{\rm ext}$ is applied along the left-right extrema. We implicitly assume the system is always in the linear regime.
The set of linear equations is written as $\hat A \vec{X} = \vec{B}$, where $\vec{X}$ contains all the unknown currents $I_x^{ij}$, $I_y^{ij}$ and voltages $V^{ij}$ on the square lattice, $\vec{B}$ contains the external voltages, either $0$ or $V_{\rm ext}$,
$\hat A$ is a sparse matrix containing, at most, three non-zero terms in each row, i.e., $\pm 1,Z(\vec{x})$.
Once the linear system is solved, the total impedance along the left-right extrema is then 
\begin{equation}
    Z_{\rm tot} = \frac{V_{\rm ext }}{\sum_{ij} I_x^{ij} } (L-1).
\end{equation}
A more detailed description of the RIN equations can be found in \citeSM{Venditti2023RIN}.

In that previous work, we only considered constant values of the normal state resistivities $R_N$ and inductances $L_i$.
To give a more accurate physical description, we here introduce Halperin-Nelson corrections on the normal-state resistivity, given by vortex-antivortex and Cooper pair renormalization effects above the critical temperature~\citeSM{halperin1979resistive}. 

We also consider an $s$-wave temperature dependence of the local inductances to show how \emph{apparent unconventional} behaviors can emerge from inhomogeneities 
\begin{equation}
    L_i=L_s g(\tau_i) \tanh\left(\frac{1.764}{2} \frac{g(\tau_i))}{\tau_i} \right),\qquad 
    \tau_i= \frac{T}{T_{c,i}}\\
\end{equation}
where we adopt $g(\tau)=\sqrt{\cos\left(\frac{\pi}{2}\tau^2\right)}$ as the interpolating formula for the BCS solution.

\begin{figure}
    \centering
    \includegraphics[width=\linewidth]{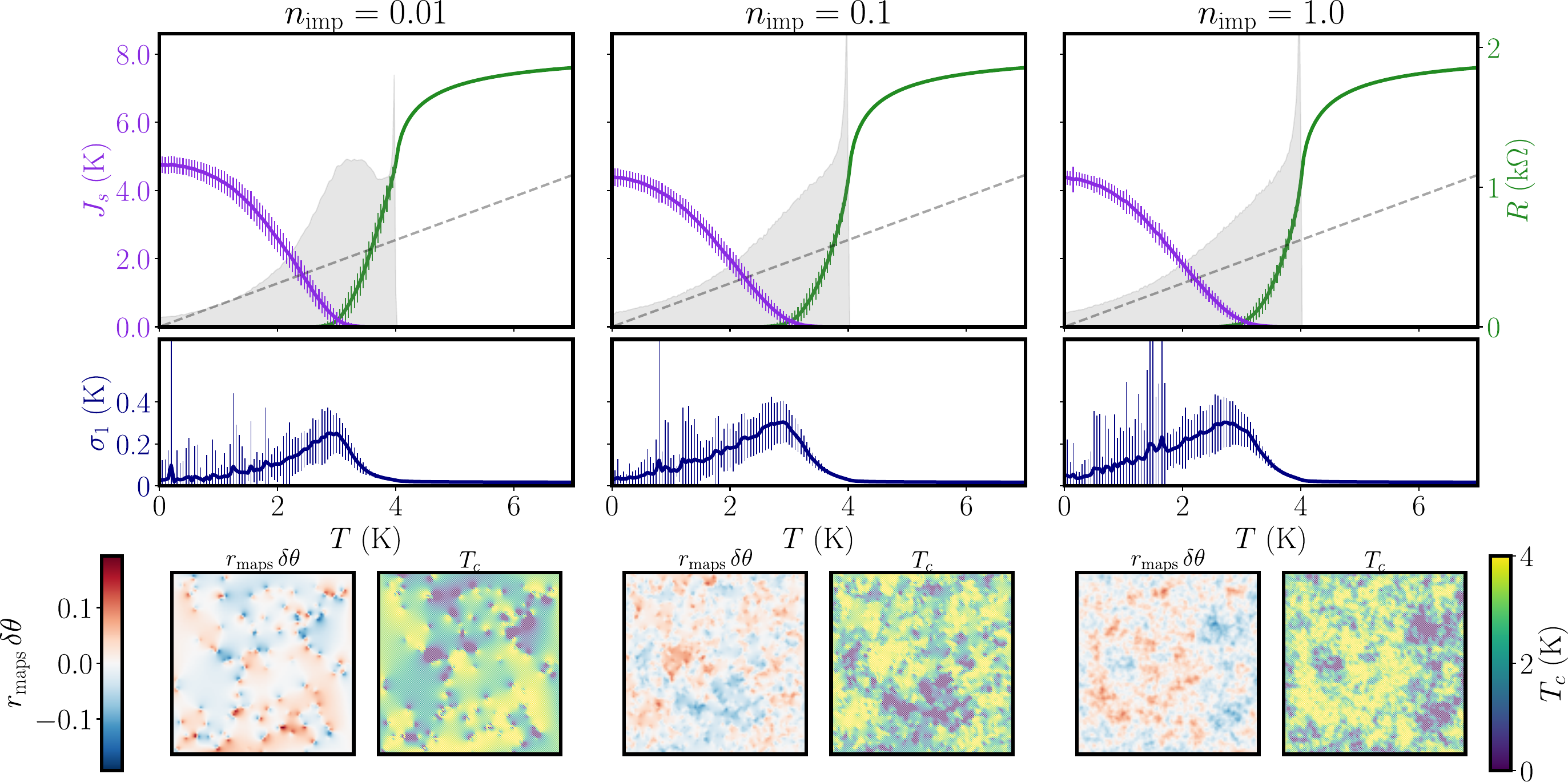}
    \caption{ 
    RIN calculations at fixed $\sqrt{{\rm Var}[\theta]}=0.03^\circ$ and $\theta_0=1.05^\circ$, for $n_{\rm imp}=0.01$ (left column), $n_{\rm imp}=0.1,$ (central column), and $n_{\rm imp}=1$ (right column). 
    (Upper panels) Superfluid stiffness $J_s$ in Kelvin (violet, left axis) and DC resistivity (green, right axis);
    (central panels) real part of the conductivity $\sigma_1$ in Kelvin;
    (lower panels) configurations examples of the $r_{\rm maps}\delta\theta$ maps and their corresponding local critical temperatures.
    Error bars refer to 100 disorder configurations. 
    }
    \label{SM-fig:transport-vs-nimp}
\end{figure}

The transport properties are tied to the average value $\theta_0$ and variance of the twist angle distribution Var$[\theta]$, this last one being the relevant observable parameter, which we tune by means of $r_{\rm maps}$. 
Once the amplitude of the twist variations is rescaled and the variance is fixed, the exact value of $n_{\rm imp}$ becomes irrelevant, as it can be seen in Fig.~\ref{SM-fig:transport-vs-nimp}, where we show the AC and DC transport, the induced $P(T_c)$, as well as examples of possible configurations for $n_{\rm imp}=0.01,\,0.1,\, 1$.
As one can see, there are no significant differences in their transport properties.
We therefore fix $n_{\rm imp}=0.1$ in the main text and following discussion, which already show qualitative resemblance with observed twist angle inhomogeneities in twisted bi- and tri-layer graphene~\citeSM{Uri2020,Turkel2022,Beechem2014,Hu.2024.CPL}.

Figure~\ref{SM-fig:suppl_all-configs} presents the full set of calculations we did for all parameters Var[$\theta$], $\theta_0$, and for all configurations.
Notice that, in an experimental setup, the exact realization of twist angles in the device can give rise to very different behaviors in the stiffness and the optical conductivity, which result in the large errorbars displayed in the main text.

\begin{figure}
    \centering
    \includegraphics[width=0.85\linewidth]{SM-Fig6.pdf}
    \caption{
    Exact RIN solutions of the DC resistivity (green, right axis), and $\sigma_1,\,\sigma_2$ (respectively in black and purple, left axis), for all 100 configurations of disorder ($n_{\rm imp}=0.1$) and ranging parameters ($\theta_0=1^\circ,1.02^\circ,1.05^\circ, 1.1^\circ$, $\sqrt{{\rm Var}(\theta)}=0.008^\circ,0.015^\circ, 0.030^\circ, 0.076^\circ$), labeled in each of the panels as $(\theta_0\pm \sqrt{{\rm Var}[\theta]})^\circ$.
    The values for local resistors and local inductances are taken as
    $R_N=2\,{\rm k\Omega}$, $L_m=$10\,nH $L_s=$1\,nH.
    }
    \label{SM-fig:suppl_all-configs}
\end{figure}

\subsection{Comparison with EMT}
In Fig.~\ref{SM-fig:suppl_RIN-vs-EMT} we present the comparison between RIN and EMT for the whole set of parameters.
The more $P(\theta)$ gets away from $\theta_{\rm MA}$ and gets broadened, the less the EMT is capable of capturing the exact RIN solution.
The vertical line indicates the percolation threshold of the EMT where $w_{\rm sp}=0.5$.
When $\theta_0=1^\circ$ (upper panels), the system never reaches $w_{\rm sp}$ and $R>0$ at all temperatures, both for the EMT solution and the {\rm average solution} of the RIN. 
Notice again that some specific realization of disorder can result in a truly superconducting state with $R=0$ at a finite temperature. 
The AC response is also sensitive to the presence of superconducting puddles, and hence $\sigma_2$ can be finite even when the system doesn't reach a proper superconducting state (Fig.~\ref{SM-fig:suppl_RIN-vs-EMT} upper panels).
In such situations, the EMT underestimates the AC response while also not capturing the long-tailed behavior of the DC resistivity.
On the contrary, in intermediate situations of correlated disorder (see in Fig.~\ref{SM-fig:suppl_RIN-vs-EMT} the panels with $\theta_0=1.02^\circ,\, 1.05^\circ$ and also $\theta_0=1.1^\circ$, $\sqrt{{\rm Var}[\theta]}=0.076^\circ$) the EMT solution overestimates the superfluid stiffness.
Furthermore, the EMT percolating temperature does not always coincide with the onset of a finite $J_s$.
\begin{figure}[h]
    \centering
\includegraphics[width=0.98\linewidth]{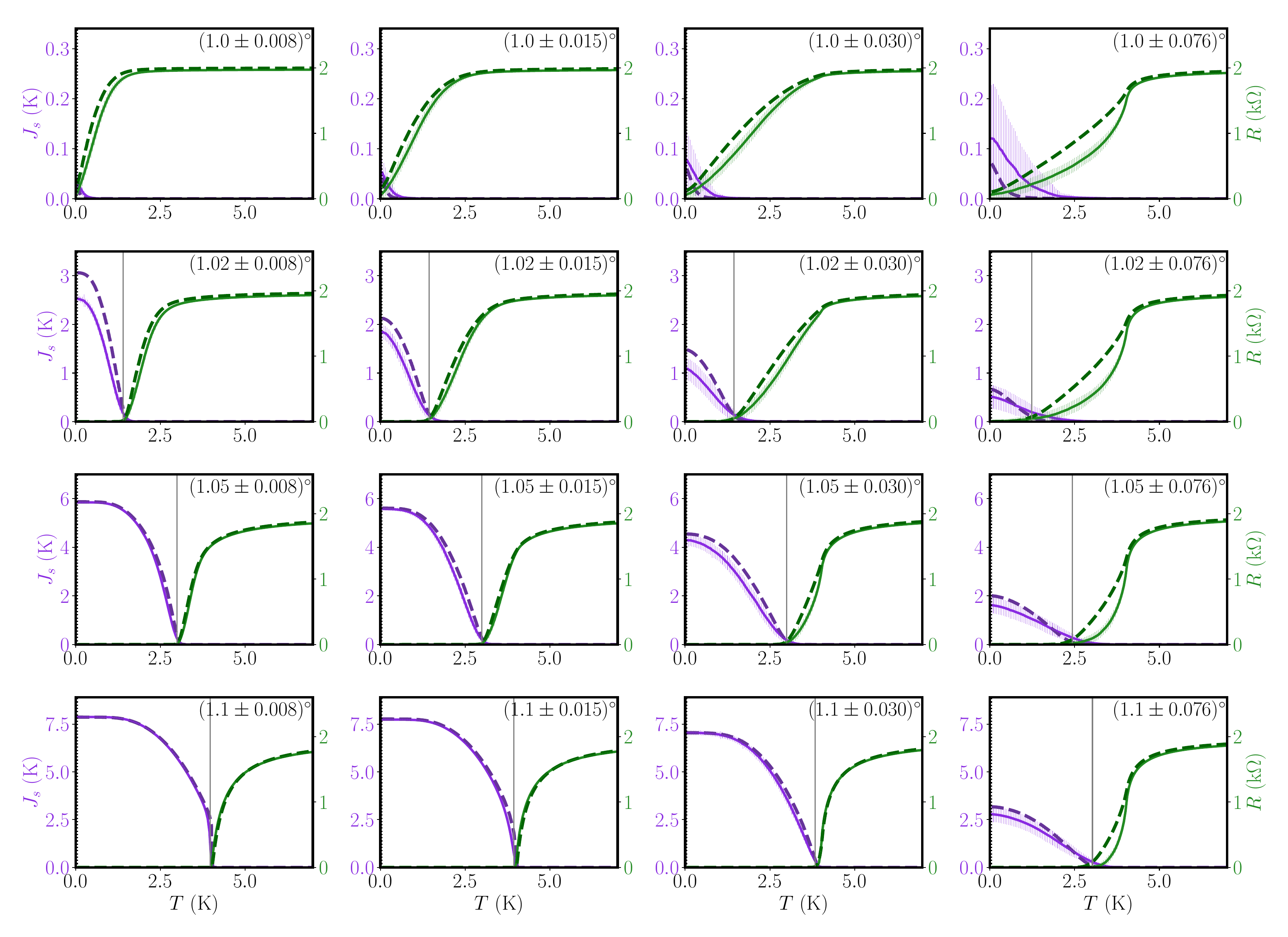}
    \caption{
    Exact RIN calculations (full lines with errorbars) compared to the EMT approximate solution (dashed lines) of the superfluid stiffness $J_s$ (purple, left axis) and DC resistivity (green, right axis) for the whole set of parameters ranging in $\theta_0=1^\circ,1.02^\circ,1.05^\circ, 1.1^\circ$, and $\sqrt{{\rm Var}(\theta)}=0.008^\circ,0.015^\circ, 0.030^\circ, 0.076^\circ$), labeled in each of the panels as $(\theta_0\pm \sqrt{{\rm Var}[\theta]})^\circ$.
    The values for local resistors and local inductances are taken as
    $R_N=2\,{\rm k\Omega}$, $L_m=$10\,nH $L_s=$1\,nH.
    The vertical line indicates the EMT percolation threshold.}
    \label{SM-fig:suppl_RIN-vs-EMT}
\end{figure}

\begin{figure}
    \centering
    \includegraphics[width=\linewidth]{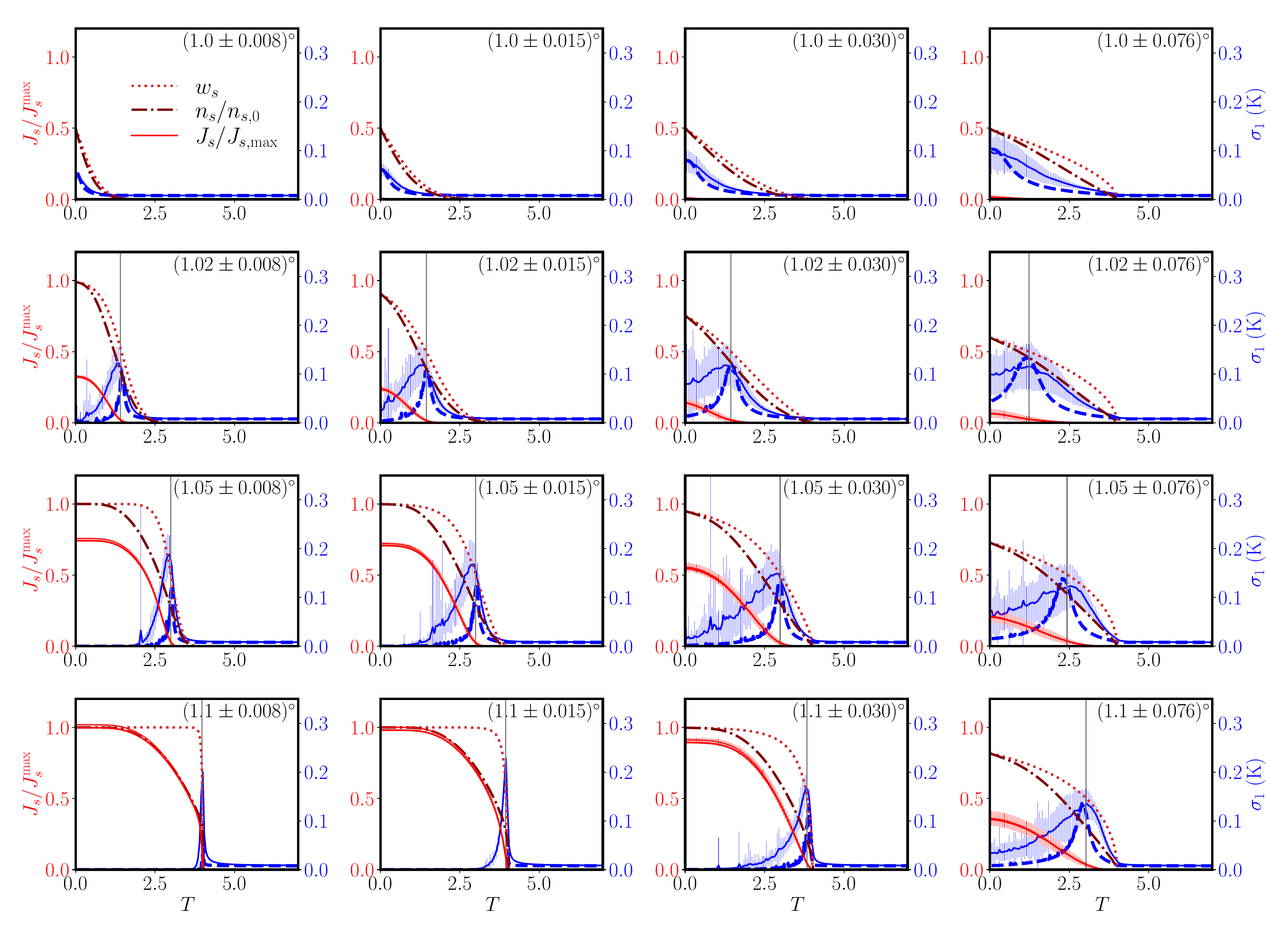}
    \caption{
    Left axis: comparison between the superfluid stiffness $J_s$ (full lines), density $n_s$ (dash-dotted line), and fraction of superconducting bonds $w_s$ (dotted).
    Right axis: RIN calculations of $\sigma_1$ (full lines with errorbars) compared with the EMT solution (dashed).
    Parameters ranges in $\theta_0=1^\circ, 1.02^\circ, 1.05^\circ, 1.1^\circ$, and $\sqrt{{\rm Var}(\theta)}=0.008^\circ,0.015^\circ, 0.030^\circ, 0.076^\circ$).
    The vertical line indicates the EMT percolation threshold $T_p$ defined as $w_{\rm sc}P(T_p)=w_{\rm sp}=0.5$.
    }
    \label{SM-fig:suppl_Js-vs-ns_sigma1}
\end{figure}

Finally, let us underline the difference between {\em global} superfluid stiffness $J_s$ and the spatially averaged local superfluid density $n_s$.
In homogeneous systems, in fact, $J_s\propto n_s$, whereas such proportionality does not hold in inhomogeneous systems. 
In Fig.~\ref{SM-fig:suppl_Js-vs-ns_sigma1} (left axis) we present our data for $J_s$ (full lines), compared with $n_s$ (dash-dotted) and $w_s$ (dotted). We recall that $w_\mathrm{\rm sc}(T)=w_{\mathrm{sc},0} \int_T^\infty P(T_c)dT_c$ is the density of superconducting bonds, and $n_s(T)=\int_T  f_{\rm BCS}(T,T_{c,i}^0) P(T_c^i) dT_c^i$, where $P(T_c)$ are the ones presented in the lower panels of Fig.~\ref{SM-fig:PTc}.
To have a better comparison between superfluid stiffness and density, we normalize both of them with respect to the cleanest case. In particular, $J_s^{\rm max} = \frac{\hbar^2}{4e^2 L_s}$ is the zero-temperature value of the clean system. 
On the right axis, we plot the corresponding $\sigma_1$.
In the clean limit, where the EMT percolating temperature coincides with a sharp peak of $\sigma_1$, the normalized $n_s$ and $J_s$ coincide. Here, the fraction of superconducting bonds $w_s$ is almost a step function at $T_{c}\approx T_{c,\rm max}$.
As the peak of $\sigma_1$ broadens, the two quantities drift apart. 
The EMT $\sigma_1$ (dashed, right axis) is always peaked at the percolation temperature. 
In the exact RIN solution $\sigma_1$ is instead much broader, showing a substantially different functional form compared to its EMT counterpart, with an important residual value symptomatic of long-range correlations.

\clearpage
\begingroup\SMbibsetup
\endgroup

\end{document}